\documentclass{aa}
\usepackage{natbib}
\usepackage{graphicx}
\usepackage{txfonts}
\usepackage{url}

\def\PSRB1259{PSR B1259-63/LS 2883}
\def\PSRJ2032{PSR J2032+4127/MT91 213}
\def\HESSJ0632{HESS J0632+057}
\def\LSI61{LS I +61$^{\circ}$303}

\usepackage[normalem]{ulem}

\begin{document}

\title{Modelling the correlated keV/TeV light curves of Be/$\gamma$-ray binaries}

\author{A. M. Chen\inst{1} \and J. Takata\inst{1} }

\institute{
Department of Astronomy, School of Physics, Huazhong University of Science and Technology, Wuhan 430074, China. \\
chensm@mails.ccnu.edu.cn; takata@hust.edu.cn
}

\date{Received xxx / Accepted xxx}

\abstract{
Be/$\gamma$-ray binaries comprise a confirmed or presumptive pulsar orbiting a Be star and emit luminous $\gamma$-rays. Non-thermal emissions are thought to arise from synchrotron radiation and inverse-Compton (IC) scattering in the shock where the pulsar wind is terminated by the stellar outflow.
We study wind interactions and shock radiations from such systems and show that the bimodal structures observed in keV/TeV light curves are caused by enhanced synchrotron radiation and IC scattering during disc passages. We use a simple radiation model to reproduce orbital modulations of keV X-ray and TeV $\gamma$-ray flux and compare with two confirmed pulsar/Be star binaries (i.e. \PSRB1259 and \PSRJ2032), and two candidates (i.e. \HESSJ0632 and \LSI61).
We find that the keV/TeV light curves of former two binaries can be well explained by the inclined disc model, while modelling the modulated emissions of latter two sources remains challenging with current orbital solutions. Therefore, we propose alternative orbital geometries for \HESSJ0632 and \LSI61. We estimate the positions and inclination angles of Be discs by fitting correlated keV/TeV light curves. Our results could be beneficial for future measurements of orbital parameters and searches for radio pulsations from presumed pulsars.
}

\authorrunning{Chen \& Takata}
\titlerunning{The keV/TeV emissions from Be/$\gamma$-ray binaries}

\maketitle

\section{Introduction}
$\gamma$-ray binaries are a rare subclass of high-mass binary systems harbouring a compact object in orbit with a type O/B star and radiate luminous $\gamma$-rays with non-thermal energy spectra peaking beyond 1 MeV (see Dubus 2013, 2015; Lamberts 2016; van Soelen 2017; Paredes \& Bordas 2019; Chernyakova \& Malyshev 2020 for reviews).
So far, less than ten such binaries have been found, and only two of them with compact objects have been identified as pulsars. The lack of accretion emission in other binaries indicates that the unknown compact objects are likely to be non-accreting neutron stars (NSs), although stellar-mass black holes cannot be completely ruled out (Dubus 2006a).
Depending on the spectral types of massive companions and the presence of a decretion disc, the detected $\gamma$-ray binaries can be divided into the following two types:
\begin{itemize}
  \item O stars: such as LS 5039 (Motch et al. 1997; Aharonian et al. 2005a),  1FGL J1018.6-5856 (Fermi LAT Collaboration 2012; H.E.S.S. Collaboration 2015a), LMC P3 (Corbet et al. 2016; H.E.S.S. Collaboration 2018), HESS J1832-093 (H.E.S.S. Collaboration 2015b; Eger et al. 2016), and 4FGL J1405.1-6119 (Corbet et al. 2019);
  \item Be stars: such as PSR B1259-63/LS 2883\footnote{At first, the massive star, LS 2883, was identified as a B2e-type star by Johnston et al. (1994), but it was revised as type O9.5Ve by Negueruela et al. (2011). For simplicity, we still attribute \PSRB1259 as a Be/$\gamma$-ray binary.} (Johnston et al. 1992; Aharonian et al. 2005b), PSR J2032+4127/MT91 213 (Lyne et al. 2015; Abeysekara et al. 2018), HESS J0632+057/MWC 148 (Aharonian et al. 2007; Hinton et al. 2009), and LS I +61$^{\circ}$303 (Gregory \& Taylor 1978; Albert et al. 2006).
\end{itemize}
Both types show similar spectral characteristics with orbital modulated radiations from radio to $\gamma$-rays, suggesting that they are likely powered by spin-down of pulsars, as in \PSRB1259. However, there are significant differences between these two types. For example, those $\gamma$-ray binaries hosting O stars usually have shorter orbital periods and lower eccentricities than those with Be companions (hereafter Be/$\gamma$-ray binaries). The former type usually exhibits a single-peak profile in its light curves, which is typically attributed to Doppler-boosted shock emission around inferior conjunction (INFC, Dubus et al. 2010; Takata et al. 2014; Molina \& Bosch-Ramon 2020). The keV/TeV light curves of  Be/$\gamma$-ray binaries usually display bimodal structures, which are likely caused by the compact objects interacting with the discs of their massive companions (Tavani \& Arons 1997; Chen et al. 2019).

Among detected Be/$\gamma$-ray binaries, \PSRB1259 has been thoroughly investigated  both observationally and theoretically. The most notable feature of this binary is the two-peak structures displayed in its keV and TeV light curves. The double-hump behaviours associated with the disappearance of radio pulsation around periastron are attributed to the pulsar crossing the inclined Be disc (Chernyakova et al. 2006, 2014; Chen et al. 2019, 2021a). The widely accepted scenario for \PSRB1259 involves a pulsar wind shock terminated by the stellar outflow, where non-thermal emissions are produced by the shock-accelerated particles (Kirk et al. 1999; Takata \& Taam 2009; Kong et al. 2011; Takata et al. 2012). Chen et al. (2019) proposed that the energy densities of magnetic field and photon field would be enhanced during disc passages, increasing synchrotron and IC luminosities, and resulting in the double peaks seen in the keV and TeV light curves.
Analogous behaviours are also seen in another pulsar/Be binary, \PSRJ2032 (Chernyakova et al. 2020b). The similar nature of these two binaries suggests that a common physical mechanism (i.e. the inclined disc model) is causing the orbital modulations of the multi-wavelength emissions.

The correlations and double-hump behaviours in the keV and TeV flux have been found in another two Be/$\gamma$-ray binaries (i.e. \HESSJ0632 and \LSI61). However, the unknown natures of their compact objects and the lack of well-measured orbital parameters make it challenging to understand the origins and orbital modulations of their high-energy radiation.
Several emission models have been proposed, including:
(1) the microquasar model, where a pair of relativistic jets produce the non-thermal emission (e.g. Bosch-Ramon \& Paredes 2004; Romero et al. 2007; Zhang et al. 2009; Massi \& Torricelli-Ciamponi 2014);
(2) the flip-flop scenario, where the compact object is an NS that evolves from a rotational regime into a propeller state along with the orbital motion (e.g. Torres et al. 2012; Papitto et al. 2012);
(3) the pulsar model, which involves a termination shock formed by collisions between the pulsar wind and the  stellar outflow as mentioned above (e.g. Cerutti et al. 2008; Sierpowska \& Torres 2009; Zdziarski et al. 2010; Zabalza et al. 2011a; Bosch-Ramon et al. 2017; Malyshev et al. 2019).

Bimodal structures are seen in the keV/TeV light curves of all four discovered Be/$\gamma$-ray binaries. In this paper, we attempt to study the origin of their correlated keV/TeV emissions following the pulsar wind interaction model and to investigate the formation of double-peak behaviours. The bimodal structures of light curves could also provide indications as to the positions and inclinations of Be discs in binaries, which are essential for measuring the orbital parameters and searching for radio pulsations from the presumptive pulsars.
The paper is organised as follows: we describe the stellar outflow of Be stars in Section 2.1; the termination shock and related radiation processes are presented in Section 2.2 and 2.3, respectively. We then display our calculated light curves with comparisons of observational data in Section 3. Finally, a brief discussion and conclusions are presented in Section 4.

\section{Model description}

\subsection{The stellar outflow}
\begin{figure}
  \centering
  \includegraphics[width=0.485\textwidth]{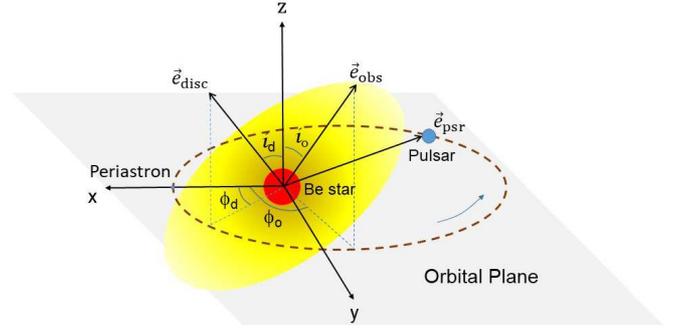}
  \caption{Illustration of the binary geometry with a pulsar orbiting around a Be star.} \label{fig:orbit}
\end{figure}


Massive Be stars are characterised by Balmer emission lines and infrared (IR) excess in their spectra (Porter \& Rivinius 2003; Rivinius et al. 2013). The origins of emission lines and IR excesses are attributed to a dense equatorial disc expelled from the fast-rotating star.
For Be/$\gamma$-ray binaries, the stellar spin axis is usually misaligned with the orbital axis, and therefore the equatorial disc is tilted relative to the orbital plane.
In Fig. \ref{fig:orbit}, we illustrate the binary geometry of a pulsar in orbit with a Be companion. We set the origin of Cartesian coordinates on the binary barycentre, the $x$-axis towards periastron, and the $z$-axis along the orbital axis. Therefore, the unit vectors in the directions of the pulsar, the observer, and the disc normal can be written as follows (Chen et al. 2021a):
\begin{eqnarray}\label{vecs}
  \vec{e}_{\rm{psr}} &=& \left( \cos \phi ,\sin \phi ,0 \right), \\
  \vec{e}_{\rm{obs}} &=& \left( \sin i_{\rm{o}}\cos \phi _{\rm{o}},
  \sin i_{\rm{o}}\sin \phi _{\rm{o}}, \cos i_{\rm{o}} \right),\\
  \vec{e}_{\rm{disc}} &=& \left( \sin i_{\rm{d}}\cos \phi _{\rm{d}}, \sin i_{\rm{d}}\sin \phi _{\rm{d}}, \cos i_{\rm{d}} \right),
\end{eqnarray}
with $\phi$ being the pulsar's true anomaly, and $i_{\rm o}$ ($i_{\rm d}$) and $\phi_{\rm o}$ ($\phi_{\rm d}$) being the inclination angle and true anomaly of the observer (the disc normal) projected on the orbital plane, respectively.

\begin{figure}
  \centering
  \includegraphics[width=0.435\textwidth]{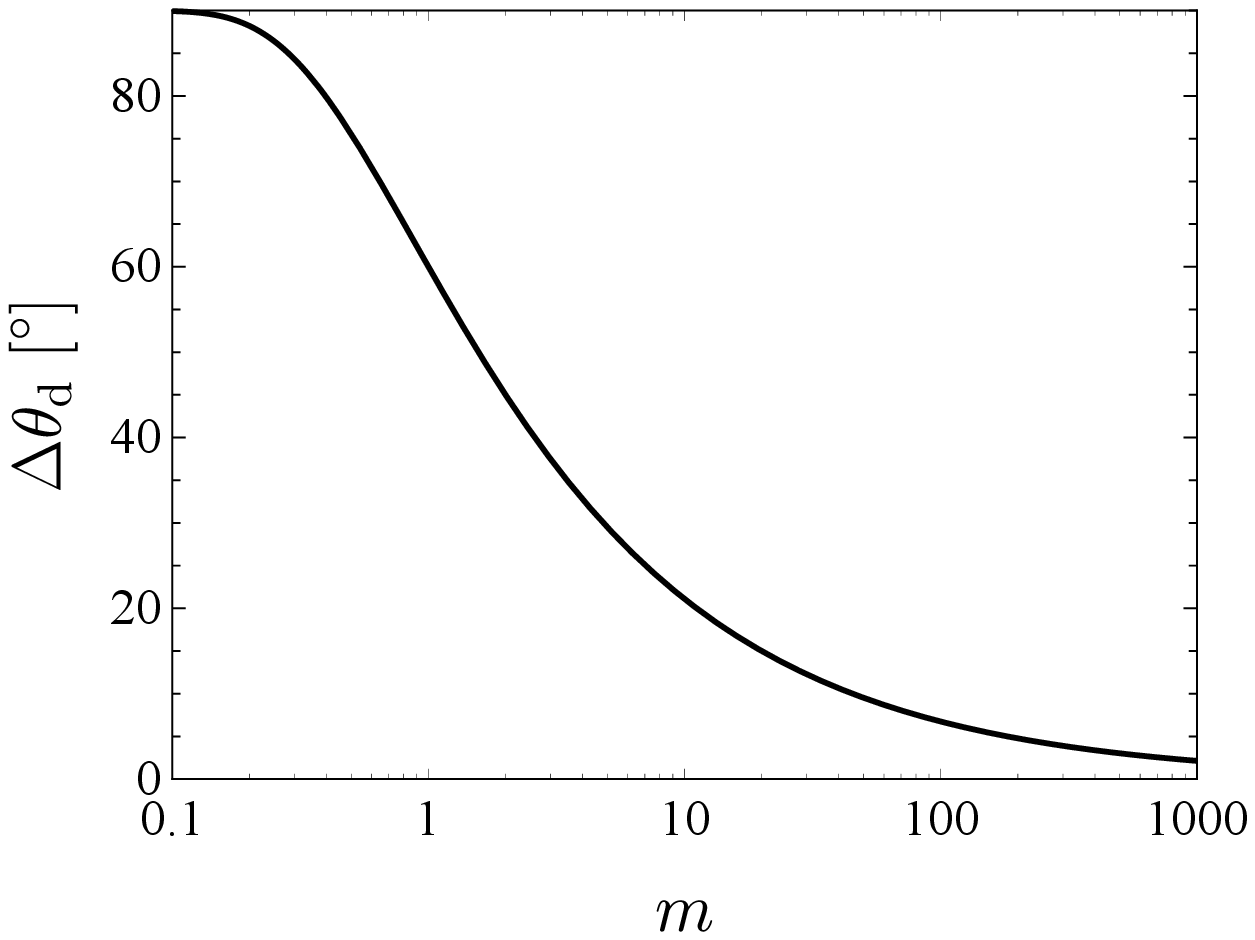}
  \includegraphics[width=0.405\textwidth]{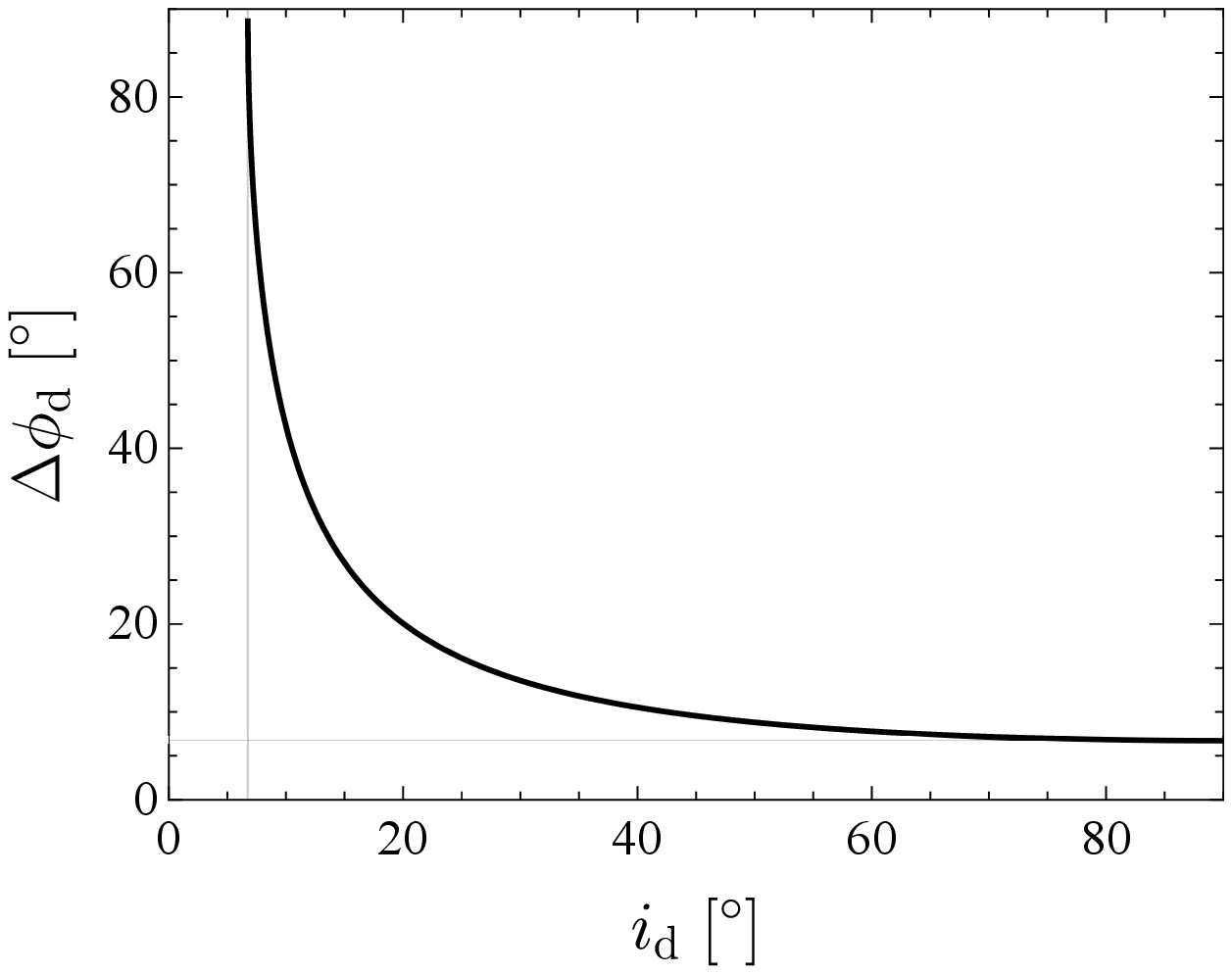}
  \caption{Top: Half-opening angle of the stellar disc $\Delta\theta_{\rm d}$ as a function of the confinement parameter $m$. Bottom: Half-opening angle of the disc projected on the orbital plane $\Delta\phi_{\rm d}$ as a function of the inclination angle $i_{\rm d}$, with the vertical and horizontal lines corresponding to $\Delta\theta_{\rm d}=6.7^{\circ}$.} \label{fig:disc}
\end{figure}

Assuming that the stellar outflow is axisymmetric along the stellar spin axis, we write the ram pressure of the outflow as (Ignace $\&$ Brimeyer 2006; Petropoulou et al. 2018; Chen et al. 2019):
\begin{eqnarray}\label{Pw}
  p_{\rm w}(R,\theta)&=&p_0R^{-2}(1+G\mid\cos\theta\mid^m),
\end{eqnarray}
with $R$ being the radial distance and $\theta$ the latitude measured from the equator. The value of $p_0$ is determined by the mass-loss rate $\dot{M}$ and velocity $v_{\rm w}$ of the polar wind as $p_0=\dot{M}v_{\rm w}$, and the second term in the bracket depicts the additional pressure of the Be disc, with $G$ being the pressure contrast. The half-opening angle of the Be disc is related to the confinement parameter $m$ as (Ignace \& Brimeyer 2006):
\begin{eqnarray}\label{dtheta}
  \Delta\theta_{\rm d}&=&\arccos(2^{-1/m}),
\end{eqnarray}
and the disc half-opening angle projected on the orbital plane is
\begin{eqnarray}\label{dphi}
  \Delta\phi_{\rm d}&=&\arcsin\left(\frac{\sin\Delta\theta_{\rm d}}{\sin{i_{\rm d}}}\right).
\end{eqnarray}
A well-confined disc requires a larger  $m$, as depicted in the top panel of Fig. \ref{fig:disc}. It is widely believed that the equatorial discs of Be stars are geometrically thin, and so we simply adopt $m=100$ in calculations (which corresponds to $\Delta\theta_{\rm d}=6.7^{\circ}$), and leave $i_{\rm d}$ as a free model parameter.
For a fixed value of $\Delta\theta_{\rm d}$, the disc region projected on the orbital plane is larger with a smaller $i_{\rm d}$, as presented in the bottom panel of Fig. \ref{fig:disc}.
If $i_{\rm d}\leq\Delta\theta_{\rm d}$, the disc is immersed on the plane, and then the pulsar will always be interacting with the disc along the orbit. When $i_{\rm d}=90^{\circ}$, we simply have $\Delta\phi_{\rm d}=\Delta\theta_{\rm d}$.

\subsection{The termination shock}
The energetic pulsar in a $\gamma$-ray binary drives a relativistic wind which is terminated by stellar outflow. The position of the termination shock is governed by the balance of two winds as:
\begin{eqnarray}\label{dyn}
  \frac{L_{\rm sd}}{4\pi r_{\rm s}^2c}&=&p_0R_{\rm s}^{-2} (1+G\mid\cos\theta\mid^m),
\end{eqnarray}
where $L_{\rm sd}$ is the pulsar's spin-down luminosity, and $c$ is the speed of light. As the pulsar moves on the orbital plane, we have:
\begin{eqnarray}\label{theta}
  \theta &=&\pi /2-\arccos \left( \vec{e}_{\rm{disc}}\cdot \vec{e}_{\rm{psr}} \right) ,
\end{eqnarray}
and the disc midplane intersects on the orbital plane at true anomalies:
\begin{eqnarray}\label{midplane}
  \phi_{\rm d, \pm}&=&\phi_{\rm d}\pm \pi/2.
\end{eqnarray}
Defining the momentum rate ratio of two winds as:
\begin{eqnarray}\label{eta}
  \eta&=&\frac{L_{\rm sd}/c}{4\pi p_0(1+G\mid\cos\theta\mid^m)},
\end{eqnarray}
the distances of the stagnation point at the shock from the pulsar and the star are:
\begin{eqnarray}\label{rs}
  r_{\rm s}&=&d \frac{\eta^{1/2}}{1+\eta^{1/2}},
\end{eqnarray}
and
\begin{eqnarray}\label{Rs}
  R_{\rm s}&=&d \frac{1}{1+\eta^{1/2}},
\end{eqnarray}
respectively, with $d$ being the orbital separation.

The post-shock magnetic field is given by the Rankine-Hugoniot relations as (Kennel \& Coroniti 1984a,b):
\begin{eqnarray}\label{B}
B&=&3(1-4\sigma)\left[\frac{L_{\rm sd}\sigma}{r_{\rm s}^2c(1+\sigma)}\right]^{1/2},
\end{eqnarray}
and the corresponding magnetic energy density is
\begin{eqnarray}\label{uB}
  u_{\rm B}&=&\frac{B^2}{8\pi},\ \rm{for}\ \phi\in[0,2\pi],
\end{eqnarray}
where $\sigma$ is the magnetization parameters of the pulsar wind, which is usually less than unity in the termination shock. Assuming $\sigma$ is constant, we have $B\propto r_{\rm s}^{-1}$ and $u_{\rm B}\propto r_{\rm s}^{-2}$. As the pulsar crosses the disc, the additional disc pressure will push the shock closer to the pulsar, and therefore the post-shock magnetic field and synchrotron radiation will be enhanced. In the following, we simply adopt $\sigma=0.001$ throughout the text.

It is usually believed that the main target photons for IC scattering in $\gamma$-ray binaries are provided by the luminous massive star. The energy density of the stellar photon field at the shock is
\begin{eqnarray}\label{ustar}
  u_{\rm star}&=&\frac{L_{\rm star}}{4\pi R_{\rm s}^2c},\ \rm{for}\ \phi\notin[\phi_{\rm d, \pm}\pm\Delta\phi_{\rm d}],
\end{eqnarray}
with $L_{\rm star}$ being the stellar luminosity. The contribution of stellar photons to IC scattering is expected to peak around super-conjunction (SUPC) or periastron phases, where the scattering is most efficient or the photon field is densest.
Alternatively, the IR emission generated by the disc may also contribute to IC, but its contribution to IC is usually much less than that of stellar photons (van Soelen \& Meintjes 2011). Therefore, an additional seed photon component is required to explain the two-peak profiles of TeV flux.

Khangulyan et al. (2012) proposed that the shock heating of the disc can supply extra soft photons for the IC process and therefore increase the $\gamma$-ray luminosity. The amount of energy available for the seed photons generated by the shock heating can be estimated from the kinetic energy of the  disc (Zabalza et al. 2011b; van Soelen et al. 2012). Under the dynamic balance, the energy density of the radiation field due to the shock-heating process may be estimated with (Chen et al. 2019)
\begin{eqnarray}\label{udisc}
  u_{\rm disc}&\simeq&\frac{\xi L_{\rm sd}}{4\pi r_{\rm s}^2c},\ \rm{for}\ \phi\in[\phi_{\rm d, \pm}\pm\Delta\phi_{\rm d}],
\end{eqnarray}
where $\xi$ is the heating efficiency. In the following, we simply adopt $\xi=0.5$ throughout the text. Its maximal contribution to IC in the shock approximately coincides with the pulsar passing through the midplane of the disc.


\subsection{Radiation processes}
The correlations as seen in the keV and TeV flux of $\gamma$-ray binaries indicate a common particle population and emitting region radiating at both energy bands (Zabalza et al. 2011a). In the pulsar scenario, the termination shock compresses the magnetic field and accelerates electrons carried by the pulsar wind. Shock-accelerated electrons emit synchrotron radiation in X-rays and upscatter the soft photons to $\gamma$-rays.
In the following, we adopt a simple radiation model to minimise model parameters and simplify calculations while maintaining the ability to reproduce the observed orbital modulations of keV and TeV fluxes.

Similar to Dubus et al. (2017), we assume that the emitting electrons follow a mono-energetic distribution with a Lorentz factor of $\gamma=10^6$, given the fact that electrons of this energy dominate the keV and TeV emissions.
We should note that the realistic particle distributions in the shock could be more complicated, and in a more sophisticated model, the acceleration and cooling processes of shocked electrons should be considered. As we focus on the orbital modulations of keV and TeV flux, assuming a more complicated distribution of particles will not significantly change our results (e.g. Dubus et al. 2017; Chen et al. 2019).
Under the mono-energetic distribution, the number of emitting particles is related to the particle luminosity $L_{\rm p}$ as (Dubus et al. 2017):
\begin{eqnarray}\label{Ne}
  N_{\rm e}&=&\frac{L_{\rm p}}{\gamma m_{\rm e}c^2}\times\tau_{\rm esc},
\end{eqnarray}
where $m_{\rm e}$ is the electron mass, and $\tau_{\rm esc}\sim d/c$ is the escape timescale of electrons from the emitting region. We assume that only a fraction of spin-down luminosity goes into particle luminosity (i.e. $L_{\rm p}/L_{\rm sd}<1$). Also, we ignore the IC cooling, which will be a good approximation for $\gamma=10^6$ in Be/$\gamma$-ray binaries with larger binary separations. The
effect of radiative cooling will be important for those binaries harbouring an O star with a much more compact orbit.

The synchrotron power radiated by an electron of Lorentz factor $\gamma$ is
\begin{eqnarray}\label{Psyn}
  P_{\rm{syn}}&=&\frac{4}{3}c\sigma _{\rm T}u_{\rm B}\gamma ^2\beta ^2,
\end{eqnarray}
with the characteristic energy of
\begin{eqnarray}\label{Esyn}
  \epsilon_{\rm{syn}}&=&h\frac{q_{\rm e}B}{2\pi m_{\rm e}c}\gamma^2\sim1.15B_{-1}\gamma_{6}^2 \ {\rm keV},
\end{eqnarray}
where $h$ is Plank constant, $\sigma_{\rm T}$ is the Thompson cross-section, $q_{\rm e}$ is the charge of the electron, $\beta=\sqrt{1-\gamma^{-2}}$, and $Q_{x}=Q/10^x$ is adopted.

When the pulsar is out of the disc region (i.e. $\phi\notin[\phi_{\rm d, \pm}\pm\Delta\phi_{\rm d}]$), the seed photons for IC scattering mainly come from the black-body photons emitted by the star. As stellar photons are radiated radially, anisotropic scattering should be considered. The anisotropic IC scattering power for a single electron is (Dubus \& Cerutti 2013)
\begin{eqnarray}\label{Pics1}
  P_{\rm{IC}}&\simeq& \frac{4}{3}c\sigma _{\rm T}u_{\rm{star}}\left( 1-\beta \mu \right) \left[ \left( 1-\beta \mu \right) \gamma ^2-1 \right]F_{\rm{KN}}\left( \gamma \right),
\end{eqnarray}
where $\mu=\vec{e}_{\rm{obs}}\cdot \vec{e}_{\rm{psr}}$ is the scattering angle, and $F_{\rm KN}(\gamma)$ is the reduction factor due to Klein-Nishina effect (Moderski et al. 2005; Khangulyan et al. 2014).
During the disc passages (i.e. $\phi\in[\phi_{\rm d, \pm}\pm\Delta\phi_{\rm d}]$), seed photons are generated by shock heating of the disc. In this case, we assume that the scattering process is isotropic, and therefore
\begin{eqnarray}\label{Pics2}
  P_{\rm{IC}}&\simeq& \frac{4}{3}c\sigma _{\rm T}u_{\rm{disc}}\gamma ^2\beta ^2F_{\rm{KN}}\left(\gamma \right).
\end{eqnarray}
The characteristic energies of up-scattered photons can be estimated as
\begin{eqnarray}\label{Eics}
  \epsilon_{\rm IC}&\simeq&
  \left\{
    \begin{array}{ll}
      4\gamma ^2\epsilon _0\sim4.0\gamma_{6}^2\epsilon _{0,\rm eV}\ {\rm TeV}, & \hbox{for $\gamma \epsilon _0/m_{\rm{e}}c^2\ll 1$,} \\
      \gamma m_{\rm{e}}c^2\sim0.5\gamma_{6}\ {\rm TeV}, & \hbox{for $\gamma \epsilon _0/m_{\rm{e}}c^2\gg 1$,}
    \end{array}
  \right.
\end{eqnarray}
where $\epsilon_0$ is the characteristic energy of seed photons.

The keV X-ray luminosity due to synchrotron radiation is given by
\begin{eqnarray}\label{Lx}
  L_{\rm X}&\sim& N_{\rm e}\times P_{\rm syn},
\end{eqnarray}
and the TeV $\gamma$-ray luminosity due to IC scattering is given by
\begin{eqnarray}\label{Lg}
  L_{\rm \gamma}&\sim& N_{\rm e}\times P_{\rm IC}.
\end{eqnarray}
With a constant injection rate of emitting particles, the modulations of synchrotron and IC scattering luminosities are mainly caused by variations in the energy densities of the magnetic field and photon field at the shock. In addition, the anisotropy of IC will also
slightly affect the $\gamma$-ray flux.

\section{Results}
In this section, we compare our calculated light curves with the observational data of four detected Be/$\gamma$-ray binaries. In the model, the keV and TeV flux are produced by synchrotron radiation and IC scattering in the shock, respectively, as depicted above.

\subsection{\PSRB1259}
\begin{figure}
\resizebox{\hsize}{!}{\includegraphics{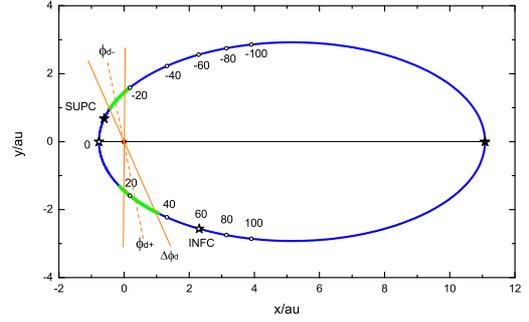}}
\caption{
Orbital geometry of \PSRB1259.
The black pentagrams are the phases of periastron and apastron, and superior and inferior conjunctions, while the empty circles mark the time intervals of every 20 d from periastron.
The positions of double peaks in keV/TeV light curves are shown in green.
The inclined disc projected on the orbital plane is illustrated in orange, with the dashed line being the midplane of the disc.
}
\label{fig:B1259_O}
\end{figure}

\begin{figure}
  \resizebox{\hsize}{!}{\includegraphics{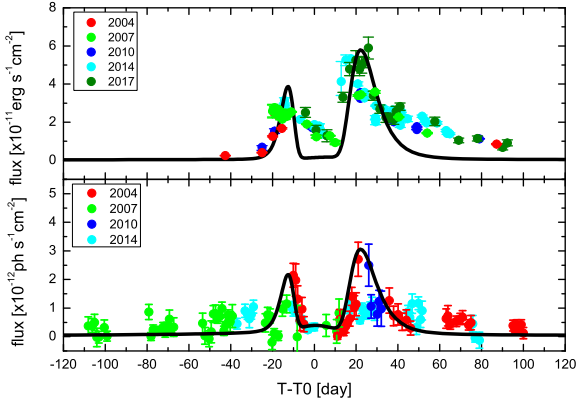}}
  \caption{
  Calculated X-ray (top) and $\gamma$-ray (bottom) light curves of \PSRB1259 with comparisons of observational data. The X-ray data during the 2004, 2007, 2010, 2014, and 2017 periastron passages are taken from Chernyakova et al. (2006, 2009, 2014, 2015) and Tam et al. (2018), respectively. The \textit{H.E.S.S.} data are from Romoli et al. (2017).
  }
  \label{fig:B1259}
\end{figure}

PSR B1259-63/LS 2883 is the first $\gamma$-ray binary where the compact object was identified as a rotational pulsar, and it is also the first variable Galactic source detected in TeV energies (Johnston et al. 1992; Aharonian et al. 2005).
The massive companion, LS 2883, with a bolometric luminosity of $L_{\star}\simeq2.2\times10^{38}\ {\rm erg\ s^{-1}}$, is characterised by an equatorial disc inclined to the orbital plane.
Long-term radio timing observations of PSR B1259-63 allow for precise measurements of related binary parameters, including the orbital period $P_{\rm o}=1236.724526\ \rm{d}$, the eccentricity $e=0.8698797$, the inclination angle $i_{\rm o}=154^{\circ}$ , and the argument of periastron $\omega_{\rm p}=138^{\circ}_.665013$ (Wang et al. 2004; Shannon et al. 2014; Miller-Jones et al. 2018). According to the parallax data of the Gaia DR2 Archive, the distance was updated to be $d_{\rm L}= 2.39\pm0.19\ {\rm kpc}$ (Gaia Collaboration 2018; Chernyakova et al. 2020a). The orbit of \PSRB1259 is illustrated in Fig. \ref{fig:B1259_O}.

Since its discovery in 1992, the binary has been intensively monitored by many telescopes, which collect a huge amount of data. Several simultaneous multi-wavelength campaigns have been performed on the system (Chernyakova et al. 2014, 2015, 2020a, 2021).
One of the most noticeable features of \PSRB1259 is the two asymmetrical peaks as seen in its keV and TeV light curves. Specifically, the X-ray flux shows a rapid increase around $T_0-15\ {\rm d}$, and reaches its maximum about $5\ {\rm d}$ later (where $T_0$ corresponds to the periastron epoch). After that, the flux decays slowly and is  then followed by another peak around $T_0+20\ {\rm d}$ (e.g. Tam et al. 2018). The $\gamma$-ray flux detected by the HESS telescope combined with all available data during the past decade also displays a similar correlated modulation as seen in the X-ray band (H.E.S.S. Collaboration 2020). The phase ranges of two peaks as seen in keV and TeV light curves are shown in Fig. \ref{fig:B1259_O} in green.

The bimodal structures of keV/TeV flux are thought to be caused by the pulsar wind interacting with stellar outflow, in particular with the inclined disc (Chen et al. 2019, and references therein). In Fig. \ref{fig:B1259}, we use the emission model described in previous sections to calculate the light curves and compare them with observational data.
The momentum rate of the polar wind adopted in the calculation is $p_0=6.3\times10^{26}\ {\rm g\ cm\ s^{-2}}$, corresponding to a typical mass-loss rate of $\dot{M}\sim 5\times10^{-8}M_{\odot}\ {\rm yr}^{-1}$ and velocity of $v_{\rm w}\sim 2\times10^{8}\ {\rm cm\ s^{-1}}$ for the polar wind of Be stars. The disc-to-wind-pressure contrast is $G=50$, which is estimated from the observed amplitudes of the peaks. As the pulsar moves into the disc region, the size of the shock shrinks because of the additional disc pressure.
As $u_{\rm B}\propto r_{\rm s}^{-2}$ and $u_{\rm disc}\propto r_{\rm s}^{-2}$, the energy densities of magnetic field and photon field are increased during disc passages. Therefore, this process will enhance synchrotron radiation and IC scattering, producing two peaks as seen in the keV and TeV light curves.
The second peak is higher than the one before periastron because of the Doppler boosting effect when the pulsar moves around INFC. The shocked flow velocity adopted in calculations is $v_{\rm s}=c/3$ (Dubus et al. 2010).
The disc parameters can be estimated by the fitting the observational data:
(1) the positions of the two peaks suggest that the midplane of the disc is located at $\phi_{\rm d,\pm}=12^{\circ}\pm 90^{\circ}$;
(2) the widths of the peaks indicate that the disc opening angle projected on the orbital plane is about $\Delta\phi_{\rm d}\simeq11.8^{\circ}$, with an inclination angle of $i_{\rm d}\simeq145^{\circ}$. This agrees with the measurement by Shannon et al. (2014), who find that the angle between the stellar axis of LS 2883 and the orbital axis is about $35^{\circ}$ (we note that PSR B1259-63 moves clockwise on its orbit as shown by Miller-Jones et al. 2018).
The position of the disc projected on the orbital plane is also illustrated in Fig. \ref{fig:B1259_O}.

We note that the observed X-ray flux decays somehow slower than the one predicted by the model. Furthermore, Chernyakova et al. (2021) discovered a third X-ray peak during the 2021 periastron passage which has never been detected before. The peak started to rise around 30 d after periastron, and did not show any clear correlated emission at other wavelengths. The slow decay behaviour and the additional peak could be due to the disc matter being piled up at the shock after disc passages, as discussed in Chen et al. (2019). Alternatively, the inhomogeneity and the clumps in the stellar outflows can also lead to the enhancement of X-rays (Chernyakova et al. 2021).

\subsection{\PSRJ2032}
\begin{table}
\caption{Three orbital solutions of \PSRJ2032 given by Ho et al. (2017). \label{tab:J2032}}
\begin{tabular}{l l l l l l l l}
\hline
Parameters&Model 1 &Model 2 &Model 3\\
\hline
$e$ & 0.936  &0.961 &0.989  \\
$P_{\rm o}(\rm{d})$ &16000 &17000 &17670  \\
$T_0$(MJD) & 58053& 58069& 58068\\
$\omega_{\rm p}(^{\circ})$ & 52& 40 &21 \\
$a\sin{i}$ (lt-s) & 7138 & 9022&16335\\
\hline
\end{tabular}
\\
{{
\footnotesize
}}
\end{table}

\begin{figure}
\resizebox{\hsize}{!}{\includegraphics{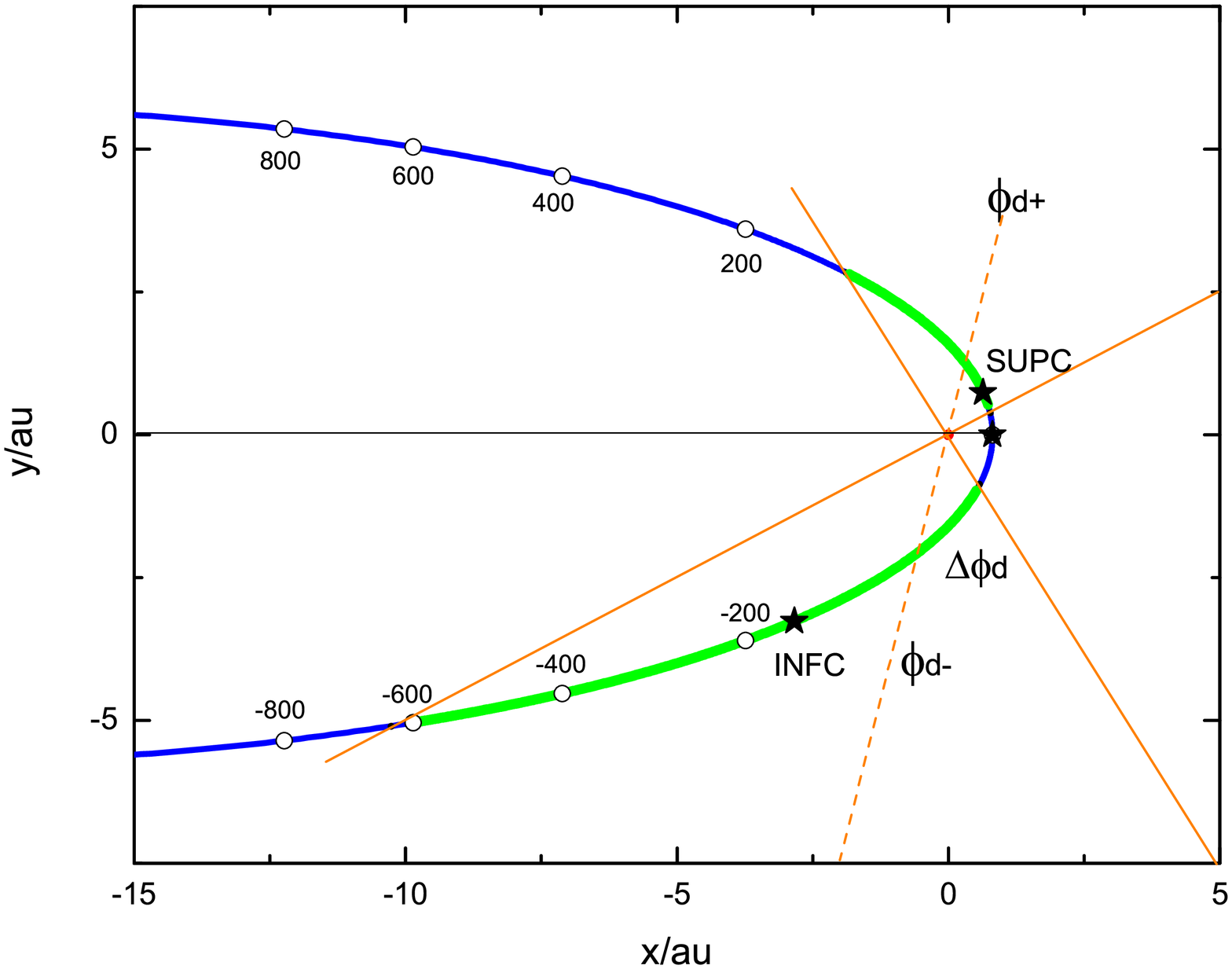}}
\caption{
Orbital geometry of \PSRJ2032 around periastron with the orbital solution of model 2 in Table  \ref{tab:J2032}.
}
\label{fig:J2032_O}
\end{figure}

\begin{figure}
\begin{minipage}{\linewidth}
  \centerline{\includegraphics[width=8.0cm]{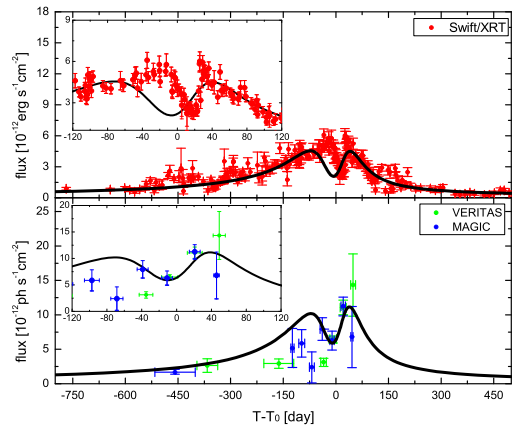}}
  \centerline{(a) Model 1}
\end{minipage}
\vfill
\begin{minipage}{\linewidth}
  \centerline{\includegraphics[width=8.0cm]{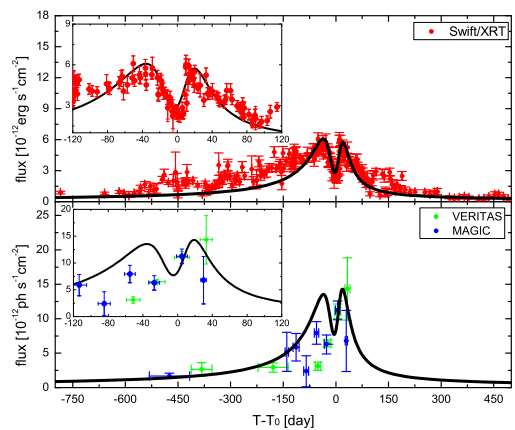}}
  \centerline{(b) Model 2}
\end{minipage}
\vfill
\begin{minipage}{\linewidth}
  \centerline{\includegraphics[width=8.0cm]{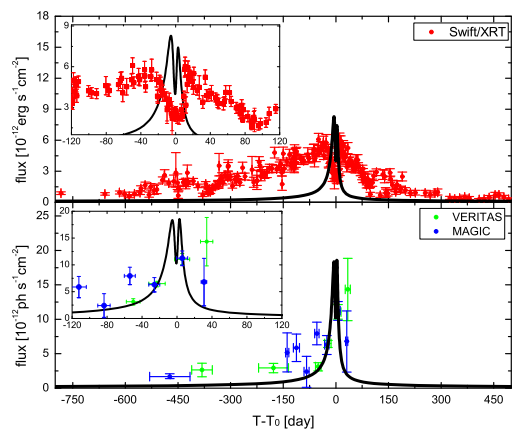}}
  \centerline{(c) Model 3}
\end{minipage}
\caption{Calculated X-ray and $\gamma$-ray light curves of \PSRJ2032 under three different orbital solutions of Ho et al. (2017). The X-ray data are analysed in this work, while the $\gamma$-ray data are taken from Abeysekara et al. (2018). The inserted plots display the light curves from between $T_0-120$ d to $T_0+120$ d.}
\label{fig:J2032}
\end{figure}

PSR J2032+4127 was firstly identified as an isolated radio-loud $\gamma$-ray pulsar with a spin period of $P=143.2\ {\rm ms}$ (Abdo et al. 2009; Camilo et al. 2009). Later radio observations found that the pulsar displays an extraordinary increase in the spin-down rate, which was attributed to the pulsar being in orbit with a massive star (Lyne et al. 2015).
The multi-wavelength monitoring conducted by Ho et al. (2017) further confirmed that PSR J2032+4127 is moving in a highly eccentric orbit ($e\sim0.94-0.99$) around the B0Ve star MT91 213 with a very long period $P_{\rm o}\sim 16000-17670\ {\rm d}$. The above characteristics suggest that \PSRJ2032 is a similar $\gamma$-ray binary to \PSRB1259.

The X-ray flux of \PSRJ2032 increased steadily before October 2017, and was then accompanied by a rapid dip around periastron (Li et al. 2017, 2018). After that, the flux was followed by another flare that lasted for several tens of days (Coe et al. 2019; Pal et al. 2019; Ng et al. 2019).
The VERITAS and MAGIC telescopes also detected TeV $\gamma$-rays from the binary (Abeysekara et al. 2018). The TeV flux was found to increase steadily before October 2017, and was then followed by a short dip at periastron. A few days later, the flux increased again with a level comparable to the pre-periastron phase.

Due to its long period and high eccentricity, the orbital parameters of \PSRJ2032 are not well measured. Ho et al. (2017) presented three different binary models, and the related orbital parameters are summarised in Table \ref{tab:J2032}. The orbit of model 2 is illustrated in Fig. \ref{fig:J2032_O}.

Given the similar nature of the two sources, we expect the emissions from \PSRJ2032 can be explained by a similar model as that explaining \PSRB1259. Therefore, we apply the inclined disc model to calculate the expected orbital modulations of keV and TeV flux with the orbital solutions provided by Ho et al. (2017). The comparisons with observational data around periastron are presented in Fig. \ref{fig:J2032}.
The X-ray data are reproduced from the \textit{Swift/XRT} website\footnote{\url{https://www.swift.ac.uk/user_objects/}} covering the time from MJD 54087 to MJD 58560, while the $\gamma$-ray data are taken from Abeysekara et al. (2018).
The model parameters adopted in calculations are the same as \PSRB1259, except for the inclination and position angle of the disc (i.e. $i_{\rm d}$ and $\phi_{\rm d}$), and the shocked flow velocity adopted here is $v_{\rm s}\sim 0.1c$.
As we can see, although the predicted light curves following model 1 agree well with the data far before periastron, the calculated flux around periastron does not match the observations. As for model 3 with a very high eccentricity ($e=0.989$), the model also fails to fit the overall observational data.
Instead, our calculation results suggest that model 2 of Ho et al. (2017) with modest values of eccentricity and period ($e=0.961, P_{\rm o}\simeq17000\ {\rm d}$) provides a reasonable fit to the data, especially around the periastron passage.
Our fitting results suggest that the position of the midplane of the disc is located at $\phi_{\rm d,\pm}=165^{\circ}\pm 90^{\circ}$ with a rather small inclination angle $i_{\rm d}\simeq 9^{\circ}$. This corresponds to a relatively large disc opening angle on the orbital plane with $\Delta\phi_{\rm d}\simeq 48.6^{\circ}$. The position of the disc is displayed in Fig. \ref{fig:J2032_O}.

The observed X-ray flux far before periastron is slightly higher than the predicted light curve of model 2. This is probably due to oversimplifications in our disc model.
It has been suggested that the disc opening angle becomes larger with growing radius (Carciofi \& Bjorkman 2006, 2008), and therefore the pulsar would interact with the disc far before periastron. This explains why the observed X-ray flux is higher than the model prediction.
Alternatively, the discrepancy could also be due to the inhomogeneity of stellar outflow or an additional wind driven by the disc.


\subsection{\HESSJ0632}
\HESSJ0632 was discovered by the survey around the Monoceros region (Aharonian et al. 2007). Its optical counterpart, MWC 148, is a B0pe-type star with a bolometric luminosity of $L_{\star}\simeq1.6\times10^{38}\ {\rm erg\ s^{-1}}$.
Although there is no direct detection of pulsed signals, the compact object of \HESSJ0632 has been widely believed to be a rotational NS (Yi \& Cheng 2017; Bosch-Ramon et al. 2017; Barkov \& Bosch-Ramon 2018; Malyshev et al. 2019).

The system shows similar spectral index and flux variability as seen in other Be/$\gamma$-ray binaries. In particular, the keV/TeV light curves are characterised by two asymmetric peaks. The primary peak around phase $0.3-0.4$ shows a sharp increase and decrease, while the second one around $0.6-0.9$ has a flat shape with a longer duration. A clear dip around phase $0.4-0.5$ was seen at both keV and TeV bands, and the X-ray light curves also exhibit a plateau around $0.1-0.2$ (Bongiorno et al. 2011; Aliu et al. 2014; Malyshev et al. 2019; Adams et al. 2021; Tokayer et al. 2021).

\begin{figure}
\begin{minipage}{\linewidth}
  \centerline{\includegraphics[width=7.25cm]{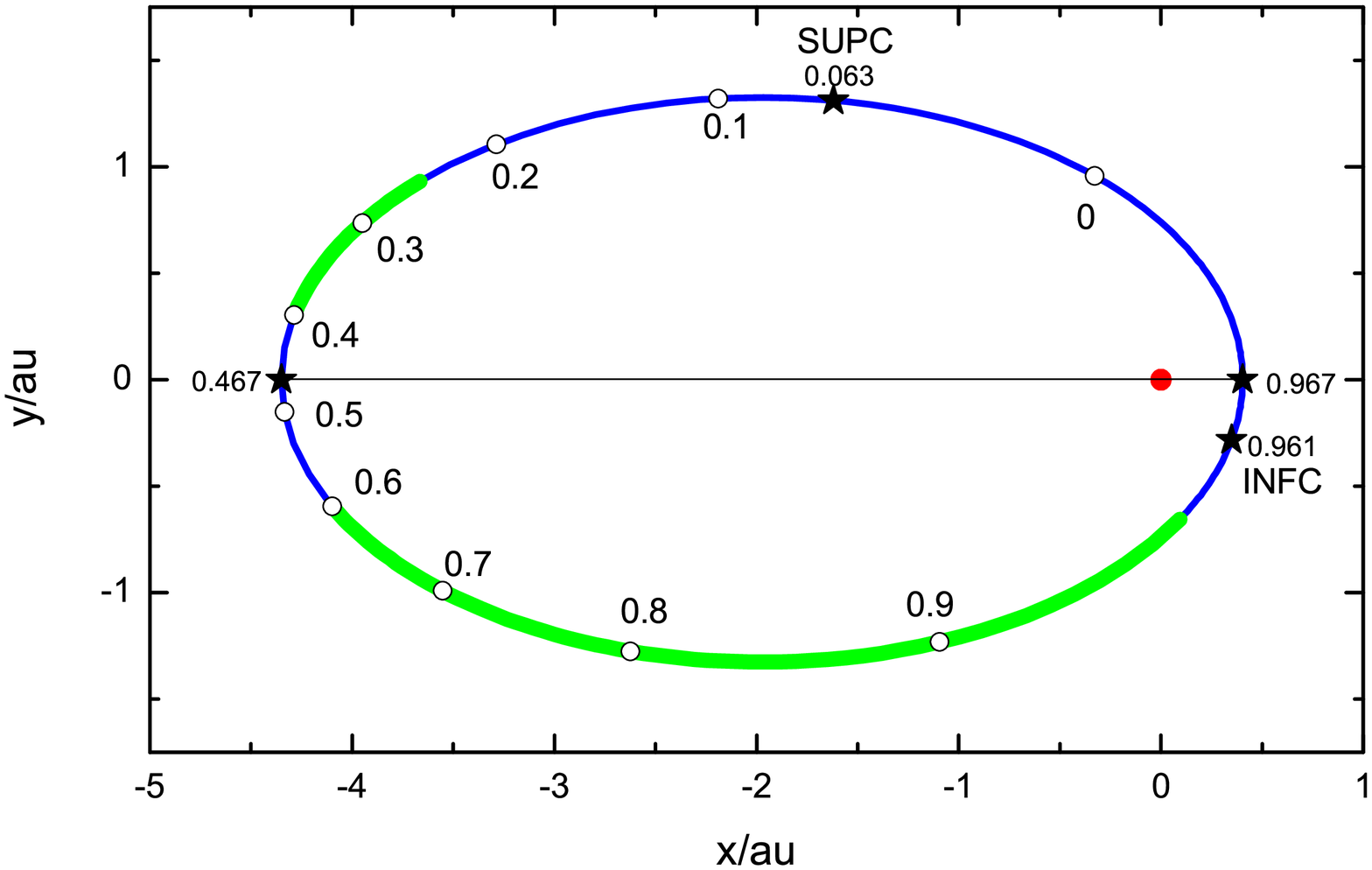}}
  \centerline{(a) Model 1}
\end{minipage}
\vfill
\begin{minipage}{\linewidth}
  \centerline{\includegraphics[width=7.25cm]{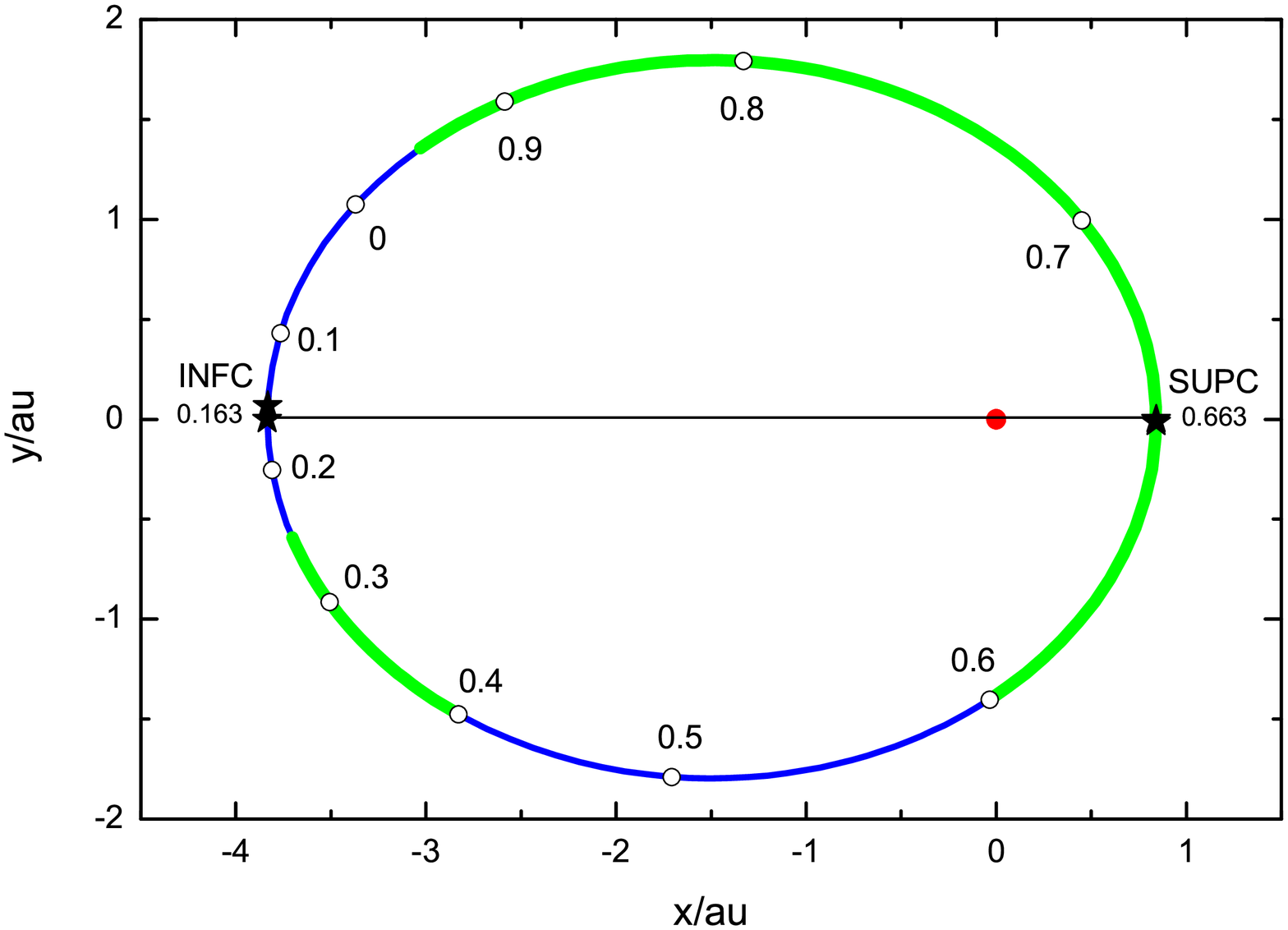}}
  \centerline{(b) Model 2}
\end{minipage}
\vfill
\begin{minipage}{\linewidth}
  \centerline{\includegraphics[width=7.25cm]{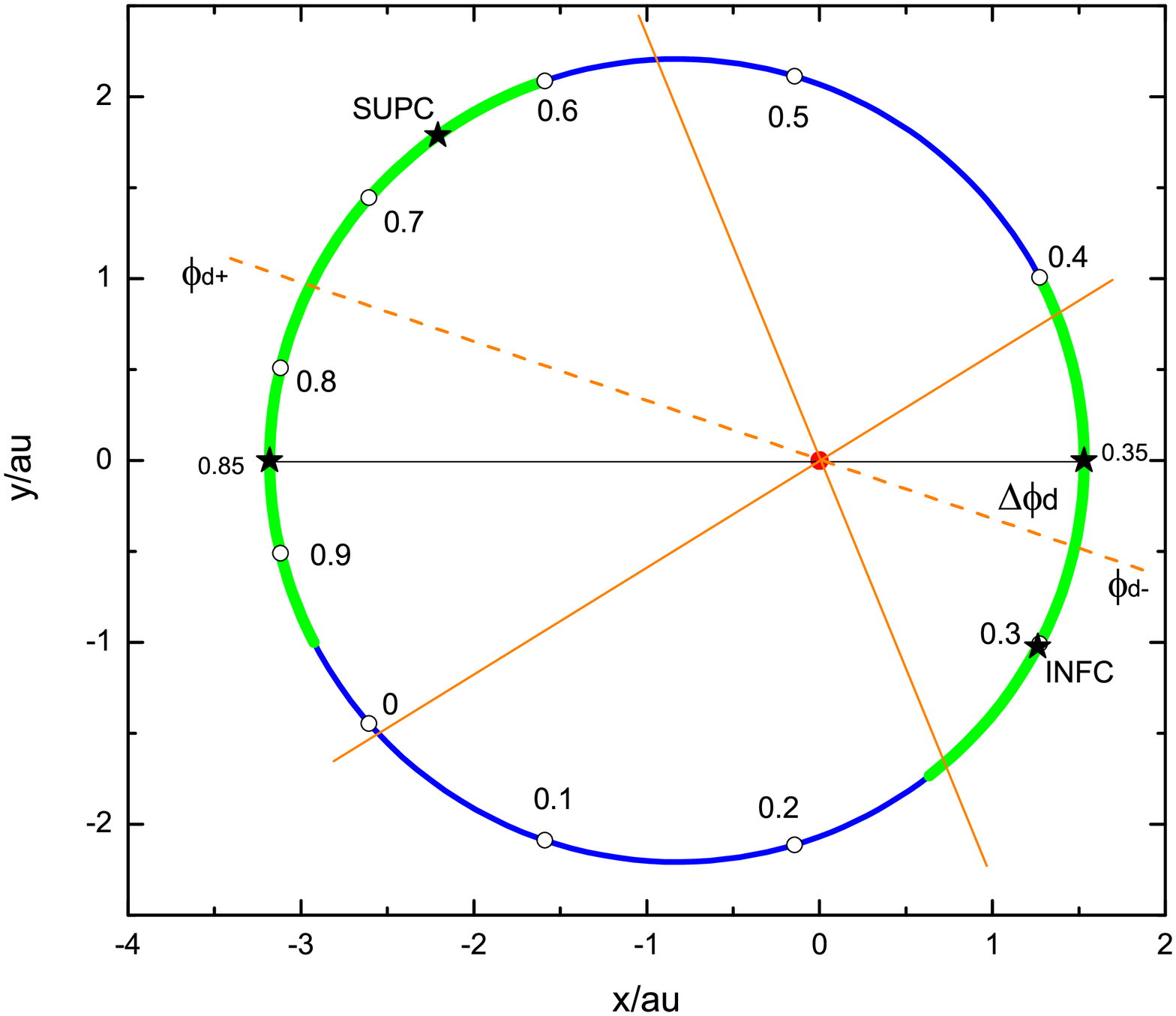}}
  \centerline{(c) Model 3}
\end{minipage}
\caption{Orbital geometries of \HESSJ0632 with three different orbital solutions given in Table \ref{tab:J0632}.}
\label{fig:J0632_O}
\end{figure}

Unfortunately, the orbital parameters of \HESSJ0632 are not
yet well measured, and different studies give different results.
Based on optical spectroscopy of the Be companion MWC 148, Casares et al. (2012) proposed an orbital period of $P_{\rm o}=321\pm5\ \rm{d}$ with an eccentricity of $e=0.83\pm0.08$ and a periastron phase of $\Phi_{\rm p}=0.967\pm0.008$. Alternatively, a slightly shorter period of $P_{\rm o}=313^{+11}_{-9}\ \rm{d}$ with a less elliptical orbit ($e=0.643\pm0.29$) was derived by Moritani et al. (2018) according to modulation of the H$\alpha$ emission line. It was also suggested that the periastron phase is around $\Phi_{\rm p}=0.663$, which is located on the opposite side of the orbit suggested by Casares et al. (2012).
Recently, Malyshev et al. (2019) and Adams et al. (2021) refined the orbital period to $317.3\pm0.7\ \rm{d}$ using the long-term X-ray observational data of \textit{Swift/XRT}, \textit{XMM-Newton}, \textit{Chandra}, \textit{NuSTAR}, and \textit{Suzaku}. Based on the positions and the relative widths of the X-ray peaks, Malyshev et al. (2019) suggested that the periastron is located around $\Phi_{\rm p}\sim0.4$, with an orbital eccentricity of $e\sim0.5$. In Table \ref{tab:J0632}, we summarise the orbital solutions of Casares et al. (2012) and Moritani et al. (2018), and the corresponding orbits are illustrated in Fig. \ref{fig:J0632_O}. The phase ranges of the two peaks as seen in keV and TeV light curves are also marked in green.

\begin{figure}
\begin{minipage}{\linewidth}
  \centerline{\includegraphics[width=7.0cm]{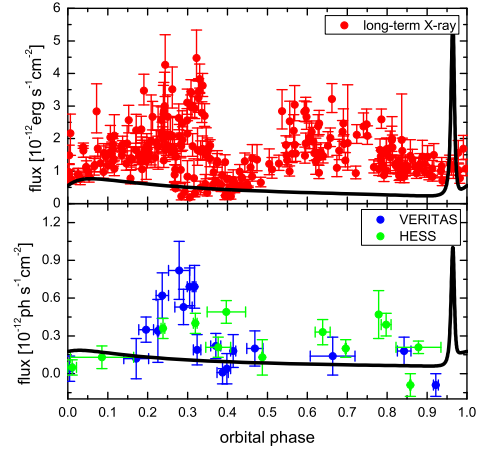}}
  \centerline{(a) Model 1}
\end{minipage}
\vfill
\begin{minipage}{\linewidth}
  \centerline{\includegraphics[width=7.0cm]{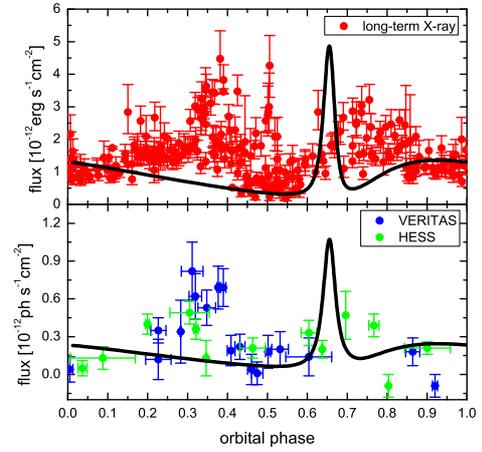}}
  \centerline{(b) Model 2}
\end{minipage}
\vfill
\begin{minipage}{\linewidth}
  \centerline{\includegraphics[width=7.0cm]{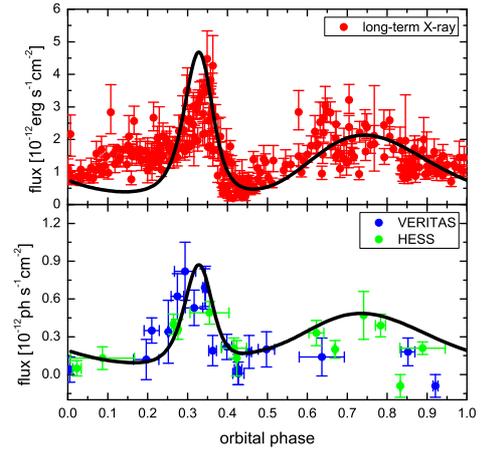}}
  \centerline{(c) Model 3}
\end{minipage}
\caption{Calculated light curves of \HESSJ0632 with comparisons of observational data under three different orbital models as given in Table \ref{tab:J0632}. The X-ray data are taken from Malyshev et al. (2019), and the $\gamma$-ray data are taken from Aliu et al. (2014). We note that the data are refolded with different orbital periods via Eq. (\ref{phase}).}
\label{fig:J0632}
\end{figure}

The similar modulations of keV and TeV flux between \HESSJ0632 and
\PSRB1259 indicate that \HESSJ0632 may also host a non-accreting NS, and the two-peak profiles are also caused by interactions between the presumptive pulsar and the inclined disc.
Therefore, in Fig. \ref{fig:J0632}, we calculate the expected shock emissions under the orbital models as given in Table \ref{tab:J0632}, and compare them with the observational data.
The X-ray data are taken from Malyshev et al. (2019), while the $\gamma$-ray data are adopted from Aliu et al. (2014). The observational data are refolded with different orbital periods in Table \ref{tab:J0632} as
\begin{eqnarray}\label{phase}
  \Phi &=&\frac{T-T_0}{P_{\mathrm{o}}}-\mathrm{int}
  \left( \frac{T-T_0}{P_{\mathrm{o}}} \right),
\end{eqnarray}
where $T_0= {\rm MJD}\ 54857.0$ is the zero phase time (Bongiorno et al. 2011).
As displayed in Fig. \ref{fig:J0632}, the expected keV and TeV emissions under the inclined disc model with the orbit solutions of Casares et al. (2012) and Moritani et al. (2018) fail to reproduce the observed modulations.
This can also be seen from the orbital phase ranges of two peaks as illustrated in the upper and middle panels of Fig. \ref{fig:J0632_O}. If the double humps are indeed caused by the presumptive pulsar passing through the disc, the true anomaly interval between two peaks should be $180^{\circ}$, and therefore even by changing the values of $i_{\rm d}$ and $\phi_{\rm d}$, it is still difficult to match the data.

Here, we propose an alternative orbital geometry for \HESSJ0632.
Because the pulsar moves faster at periastron and slower at apastron, the periastron phase should be located at the shortest separation between the peaks, around phases $0.3-0.7$ (Malyshev et al. 2019). Also, the primary peak is narrower than the second one, which puts further constraints on the periastron phase around $0.3-0.4$. Therefore, we assume that the periastron phase is around $\Phi_{\rm p}\sim 0.35$.
Also, the positions and separation of the two peaks suggest a less eccentric orbit with $e\sim0.35$. Although this eccentricity is much smaller than that of Casares et al. (2012), it is still within the error range of Moritani et al. (2018). The suggested orbital parameters are summarised as model 3 of Table \ref{tab:J0632}, and the corresponding orbit and expected light curves are shown in the bottom panels of Figs. \ref{fig:J0632_O} and \ref{fig:J0632}, respectively. The overall double-peak profiles in the keV and TeV light curves are relatively consistent with calculated results under the orbital solution of model 3, except for the X-ray plateau before the primary peak. The primary peak is significantly higher than the second one because the pulsar crosses the inner part of the disc near periastron, which pushes the shock closer to the pulsar. As $u_{\rm B}\propto r_{\rm s}^{-2}$ and $u_{\rm disc}\propto r_{\rm s}^{-2}$, enhanced synchrotron radiation and IC scattering are expected especially during the first disc passages.
Our fitting results suggest that the position of the midplane and the inclination angle of the disc are $\phi_{\rm d,\pm}=72^{\circ}\pm 90^{\circ}$ and $i_{\rm d}\simeq 12^{\circ}$, respectively. The corresponding disc opening angle on the orbital plane is $\Delta\phi_{\rm d}\simeq 34.5^{\circ}$. The derived values of the orbital parameters and the position of the disc are roughly consistent with the analysis of Malyshev et al. (2019).

It is necessary to point out that our calculated X-ray light curve does not closely match the X-ray plateau. The plateau could be caused by the clumps of the stellar outflows, as in the case of PSR B1259-63, because the clumpy outflows can push the shock closer to the pulsar side, causing a stronger magnetic field and therefore the enhanced X-ray emission (Chernyakova et al. 2021). Alternatively, the plateau could be due to the Doppler-boosting effect if the inferior conjunction is located around phase 0.1-0.2. However, the lack of a plateau phase in the TeV light curve makes this scenario unlikely.


\subsection{\LSI61}

\LSI61 was firstly discovered by the \textit{Cos B} satellite as an unknown bright $\gamma$-ray emitter (Hermsen et al. 1977), and later Gregory \& Taylor (1978) reported variable radio emissions from the exact location.
Optical spectroscopy suggested the system consists of a stellar-mass compact object in orbit with an early B0Ve-type massive star (Paredes \& Figueras 1986; Casares et al. 2005).
The nature of the compact object remains unclear even after more than 40 years of multi-wavelength observations (e.g. see Marcote 2017 and references therein).
For quite a long time, \LSI61 has been considered as a micro-quasar (e.g. Massi, Ros \& Zimmermann 2012; Massi \& Torricelli-Ciamponi 2014; Jaron 2021); however, the lack of an accretion signal in X-rays suggests that the non-thermal emission is more likely powered by spin-down of a young pulsar (Maraschi \& Treves 1981; Harrison et al. 2000; Leahy 2004; Zdziarski et al. 2010; Zabalza et al. 2011a).
Although several campaigns were conducted to search for pulsed radio and X-ray signals, no pulsations were found (Sidoli et al. 2006; McSwain et al. 2011; Ca\~nellas et al. 2012).
Recently, Weng et al. (2021) reported the detection of a transient periodic signal of 269.196 ms from \LSI61 on 2020 Jan 7 with the FAST telescope. However, no more pulsations were found in other FAST observation campaigns. This is probably caused by the strong free-free absorption by the stellar outflows, in particular with the presence of the stellar disc (Chen et al. 2021a).

According to a Bayesian analysis of the radio data, the orbital period of \LSI61 is $26.4960\pm0.0028$ d (Gregory 2002). The most peculiar feature of this system is its superorbital modulations with $P_{\rm sup}\simeq1628\pm48$ d as seen at all wavelengths, which has not been found for other $\gamma$-ray binaries (e.g., Li et al. 2011; Chernyakova et al. 2012, 2017; Ackermann et al. 2013; Ahnen et al. 2016; Massi \& Torricelli-Ciamponi 2016; Jaron et al. 2018). The origin of its long-term modulations remains unknown, although several models have been  proposed, including precession of the Be disc or the relativistic jet (Massi \& Torricelli-Ciamponi 2014; Saha et al. 2016; Xing et al. 2017; Jaron 2021).


A simultaneous observational campaign with \textit{Swift/XRT}, \textit{XMM-Newton,} and the \textit{MAGIC} telescope was conducted by Anderhub et al. (2009), which revealed a significant correlation between the keV and TeV flux. The prominent peaks around phase 0.65 are clearly seen at both energy bands, where the orbital phase is defined as Eq. (\ref{phase}) with $T_0={\rm JD}\ 2443366.775$ (Gregory 2002). Following the short dip around $0.7-0.8$, the X-ray flux displayed another slow increase from phase 0.8 to 1.0. This similar behaviour is also being found in the TeV band.

\begin{figure}
\begin{minipage}{\linewidth}
  \centerline{\includegraphics[width=6.5cm]{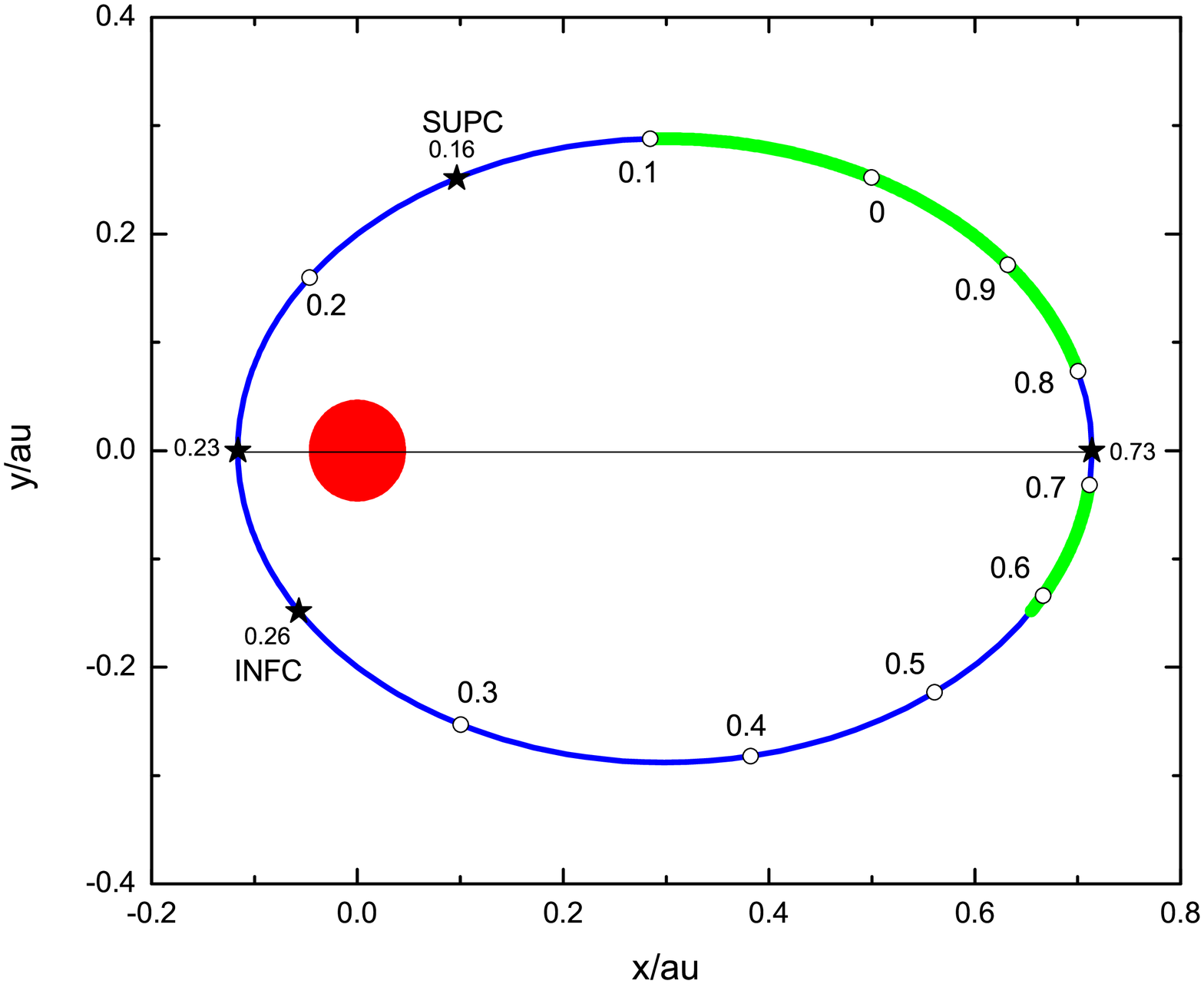}}
  \centerline{(a) Model 1}
\end{minipage}
\vfill
\begin{minipage}{\linewidth}
  \centerline{\includegraphics[width=6.5cm]{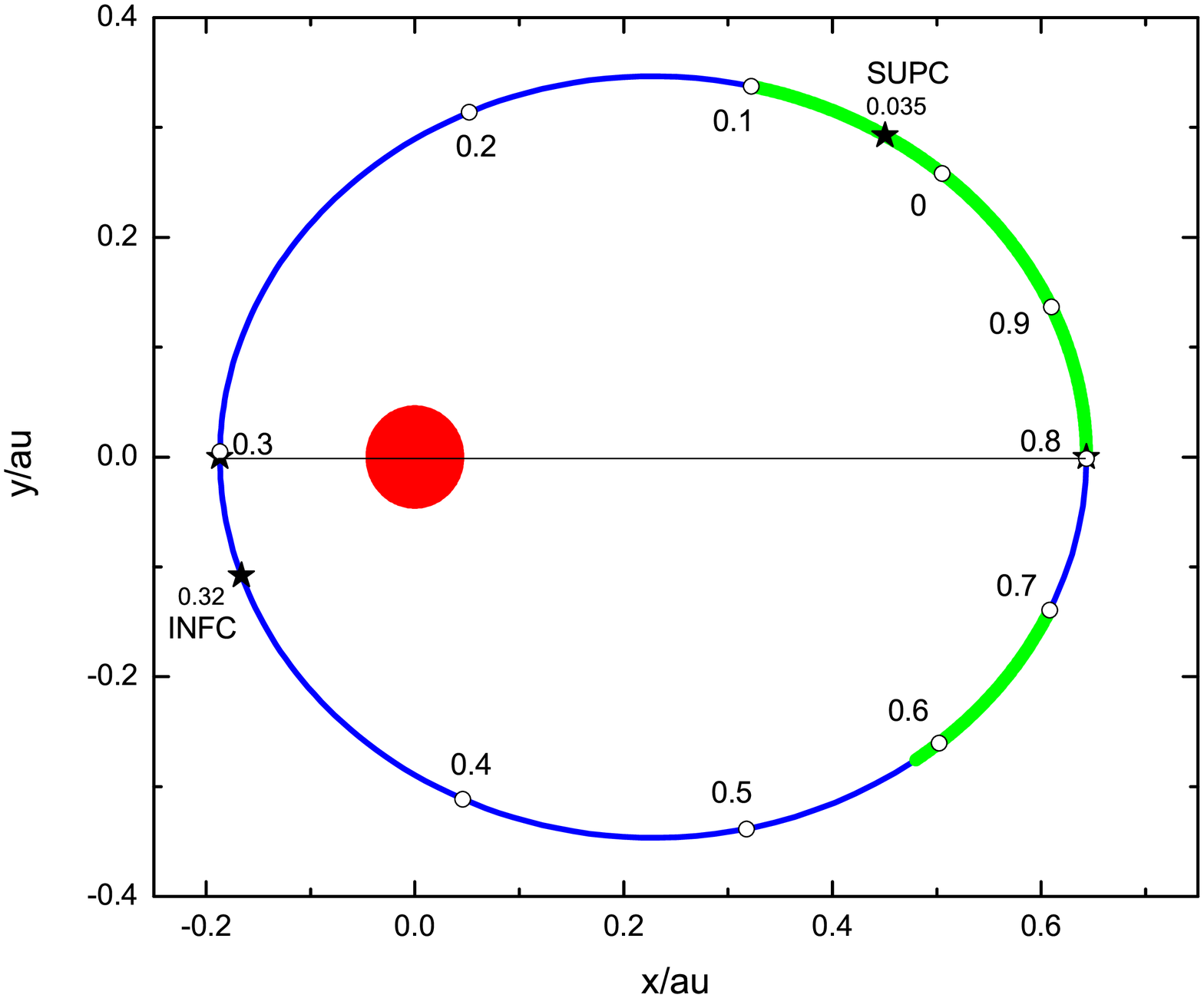}}
  \centerline{(b) Model 2}
\end{minipage}
\vfill
\begin{minipage}{\linewidth}
  \centerline{\includegraphics[width=6.5cm]{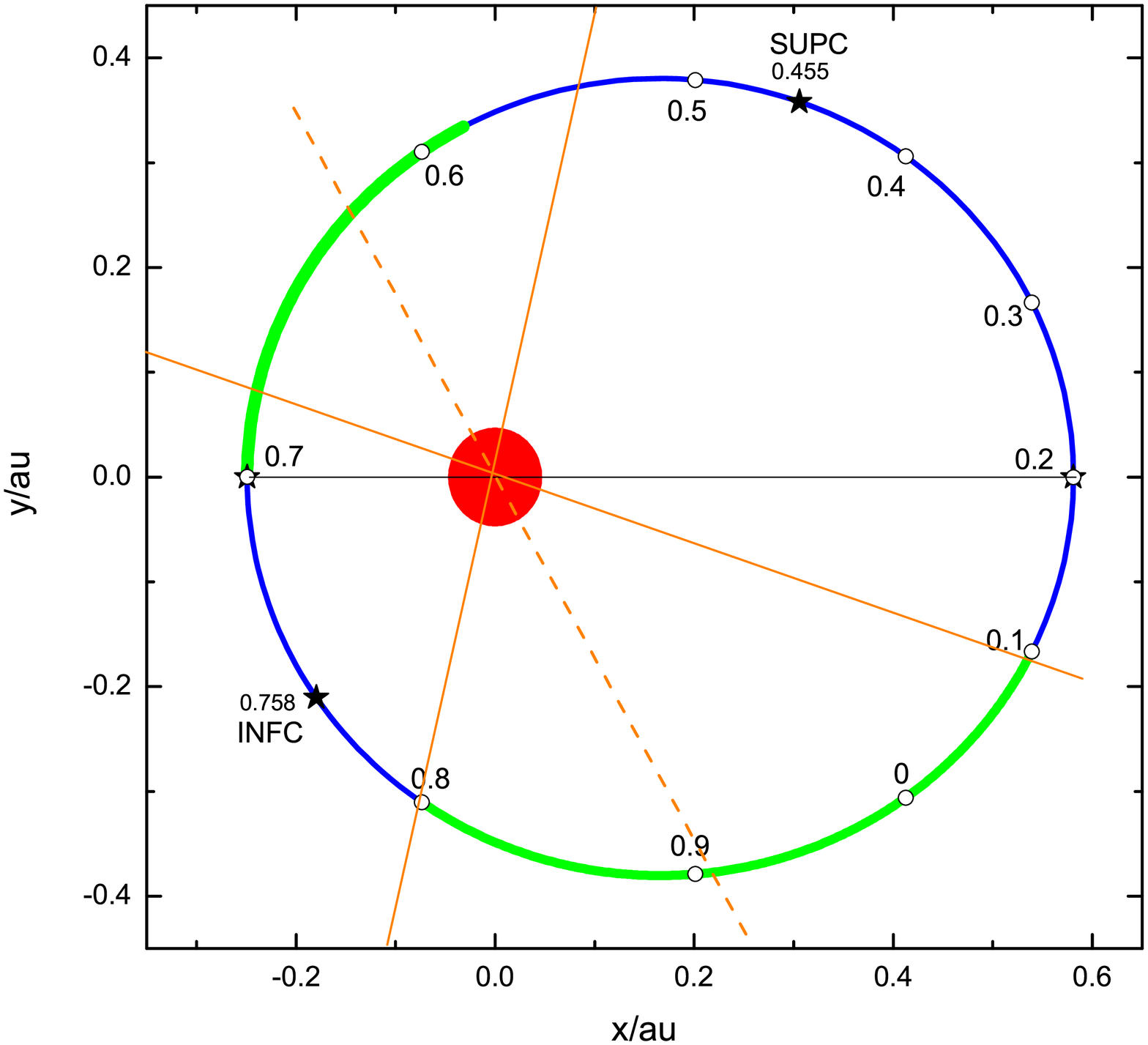}}
  \centerline{(c) Model 3}
\end{minipage}
\caption{Orbital geometries of \LSI61 with three different orbital solutions given in Table  \ref{tab:LSI}.}
\label{fig:LSI_O}
\end{figure}

Unfortunately, similar to \HESSJ0632, the binary parameters of \LSI61 are not well measured, and several orbital solutions are derived by different methods (Hutchings \& Crampton 1981; Casares et al. 2005; Grundstrom et al. 2007; Aragona et al. 2009; Kravtsov et al. 2020). The optical spectroscopy of Casares et al. (2005) suggested the eccentricity is $e=0.72\pm0.15$ with periastron occurring around phase $\Phi_{\rm p}=0.23\pm0.02$, while the H$\alpha$ monitoring campaign of Grundstrom et al. (2007) suggested $e=0.55\pm0.05$ with $\Phi_{\rm p}=0.24\pm 0.04$. More recently, a less elliptical orbit ($e<0.2$) with periastron around $\Phi_{\rm p}\sim0.6$ was derived by Kravtsov et al. (2020) based on optical linear polarisation variability curves. In Table \ref{tab:LSI}, we summarise the representative orbital models of Casares et al. (2005) and Grundstrom et al. (2007), and corresponding orbits are illustrated in Fig. \ref{fig:LSI_O}. The phase ranges of the two peaks as seen in keV and TeV bands are also marked in green.

\begin{figure}
\begin{minipage}{\linewidth}
  \centerline{\includegraphics[width=7.0cm]{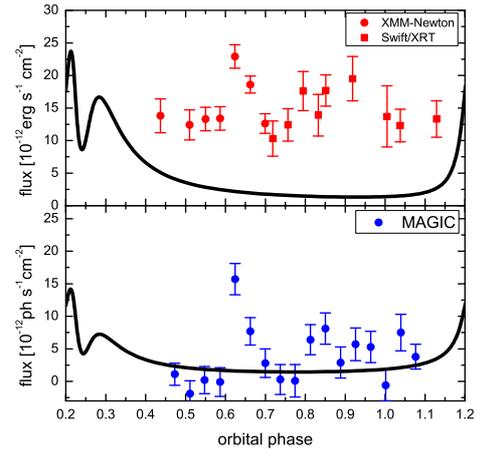}}
  \centerline{(a) Model 1}
\end{minipage}
\vfill
\begin{minipage}{\linewidth}
  \centerline{\includegraphics[width=7.0cm]{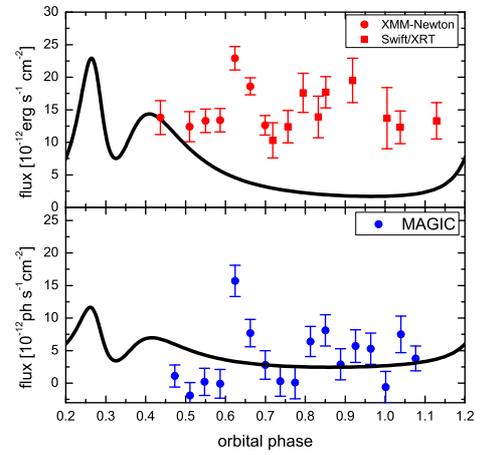}}
  \centerline{(b) Model 2}
\end{minipage}
\vfill
\begin{minipage}{\linewidth}
  \centerline{\includegraphics[width=7.0cm]{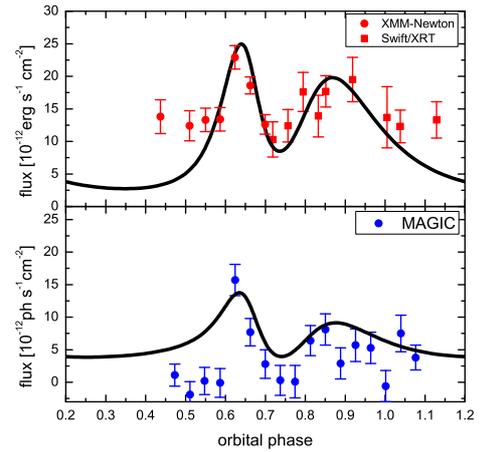}}
  \centerline{(c) Model 3}
\end{minipage}
\caption{Calculated light curves of \LSI61 with comparisons of observational data under three different orbital models as given in Table \ref{tab:LSI}. The X-ray and $\gamma$-ray data are taken from Anderhub et al. (2009).}
\label{fig:LSI}
\end{figure}

In Fig. \ref{fig:LSI}, we present the corresponding expected keV and TeV light curves under the inclined disc model with the orbital models of Table \ref{tab:LSI}. As \LSI61 shows long-term modulations in its multi-wavelength emissions, we adopt the simultaneous X-ray and $\gamma$-ray data from Anderhub et al. (2009).
As presented in the top and middle panels of Fig. \ref{fig:LSI}, the predicted light curves under the orbit solution of Casares et al. (2005) and Grundstrom et al. (2007) fail to reproduce the observed modulations of keV/TeV flux. This can also be seen from the orbital phase ranges of the two peaks as shown in Fig. \ref{fig:LSI_O}.
Similar to the analysis in \HESSJ0632, if the double-hump structures are indeed due to the pulsar--disc interactions, then the periastron phase should be located at the shortest separation between the peaks around phase $0.7-0.8$. Also, the position of the prominent peak further indicates that the periastron phase would be around $0.6-0.7$. Therefore, we assume that $\Phi_{\rm p}\sim 0.7$. This value is roughly consistent with that of Kravtsov et al. (2020). The orbital eccentricity with $e\sim0.40$ is derived from the relative width of the two peaks, and is slightly higher than that of Kravtsov et al. (2020).
The suggested orbital parameters are summarised as model 3 of Table \ref{tab:LSI}, and the corresponding orbit and expected keV/TeV light curves are shown in the bottom panels of Figs. \ref{fig:LSI_O} and \ref{fig:LSI}, respectively.
Our result suggests that the position of the midplane and the inclination angle of the disc are $\phi_{\rm d,\pm}=30^{\circ}\pm 90^{\circ}$ and $i_{\rm d}\simeq 10^{\circ}$, respectively. These correspond to the disc opening angle on the orbital plane with $\Delta\phi_{\rm d}\simeq 42.5^{\circ}$.
Similar to \HESSJ0632, \LSI61 also exhibits an X-ray plateau around phase 0.45-0.55, which could be caused by the inhomogeneity of the stellar outflows.


\begin{table}
\caption{Orbital solutions of \HESSJ0632. \label{tab:J0632}}
\begin{tabular}{l l l l l l l l}
\hline
Parameters&Model 1\dag &Model 2\ddag &Model 3\S\\
\hline
$e$ &$0.83\pm0.08$  &$0.643\pm0.29$ &0.35 \\
$P_{\rm o}(\rm{d})$ &$321\pm5$  &$313^{+11}_{-8}$   &$317.3$  \\
$\Phi_{\rm p}$  &$0.967$    &$0.663$& $0.35$\\
$\omega_{\rm p}(^{\circ})$ & $129\pm17$& $271\pm29$ &$129$ \\
\hline
\end{tabular}
\\
{{
\footnotesize
$\dag$ Casares et al. 2012;\\
$\ddag$ Moritani et al. 2018;\\
$\S$ This work. The values of the orbital period $P_{\rm o}$ and longitude of periastron $\omega_{\rm p}$ are taken from Adams et al. (2021) and Casares et al. (2009), respectively. The eccentricity and periastron phase adopted here are slightly different from that of Malyshev et al. (2019).
}}
\end{table}

\clearpage

\begin{table}
\caption{Orbital solutions of \LSI61. \label{tab:LSI}}
\begin{tabular}{l l l l l l l l}
\hline
Parameters& Model 1$\dag$  & Model 2$\ddag$ &  Model 3$\S$\\
\hline
$e$ & $0.72\pm0.15$  &$0.55\pm0.05$ & $0.40$  \\
$P_{\rm o}(\rm{d})$ & $26.496$ &$26.496$ &$26.496$  \\
$\Phi_{\rm p}$ & $0.23\pm0.02$& $0.301\pm0.011$& 0.70\\
$\omega_{\rm p}(^{\circ})$ & $21\pm13$& $57\pm9$ &$40.5$ \\
\hline
\end{tabular}
\\
{{
\footnotesize
$\dag$ Casares et al. 2005;\\
$\ddag$ Grundstrom et al. 2007;\\
$\S$ This work. The value of longitude of periastron $\omega_{\rm p}$ is taken from Aragona et al. 2009.
}}
\end{table}

\clearpage





\begin{table*}
\caption{Parameters of \PSRB1259, \PSRJ2032, \HESSJ0632, and \LSI61. \label{tab:para}}
\begin{tabular}{l l l l l l l l}
\hline
Parameters & PSR B1259-63$^a$ &PSR J2032+4127$^b$ &\HESSJ0632$^c$  &\LSI61$^d$ \\
\hline
\textit{observational parameters}\\
$e$ &  0.8698797 &0.961 &  0.35$\dag$  & 0.40$\dag$\\
$\Phi_{\rm p}$ &  - & - &  0.35$\dag$  & 0.70$\dag$\\
$P(\rm{d})$ &1236.724526 & 17000 &317.3  & 26.496\\
$d_{\rm L}(\rm{kpc})$ & 2.39 & 1.33 & 1.60  &2.30 \\
$i_{\rm o}(^{\circ})$ & 154 &$\sim60$ & $\sim60$ &$\sim45$ \\
$\phi_{\rm o}(^{\circ})$ & 132 & $\sim229$ & $\sim321$ &$\sim50$ \\
$L_{\rm sd}(\rm erg\ s^{-1})$  &  $8.2\times10^{35}$ &  $1.7\times10^{35}$ & $5\times10^{35}\dag$ &$5\times10^{35}\dag$\\
$L_{\rm star}(\rm erg\ s^{-1})$ &$2.2\times10^{38}$ &$2.6\times10^{38}$ & $1.6\times10^{38}$ &$2.6\times10^{38}$ \\

\hline
\textit{model parameters}\\
$p_0 (\rm{g\cdot cm\cdot s^{-2}})$  & $6.3\times 10^{26}$  & $6.3\times 10^{26}$ &
$2.5\times 10^{26}$ &$1.3\times 10^{26}$\\
$G$&   $50$ & 50& 50&50\\
$m$   &$100$  & 100& 100&100\\
$i_{\rm d}(^{\circ})$  &$145$  & $9$& $12$&$10$\\
$\phi_{\rm d}(^{\circ})$  &$12$  & $165$& $72$&$30$\\
$\Delta\phi_{\rm d}(^{\circ})$  &$11.8$  & $48.6$& $34.5$&$42.5$\\
$\sigma$  &$10^{-3}$  & $10^{-3}$ & $10^{-3}$&$10^{-3}$\\
$\xi$  &0.5  & 0.5& 0.5&0.5\\
$v_{\rm s}(c)$  &1/3  & 0.1& 0.02&0.1\\

\hline
\end{tabular}
\\
{{
\footnotesize
References: \\
$a$ Negueruela et al. 2011; Shannon et al. 2014; Miller-Jones et al. 2018.\\
$b$ Camilo et al. 2009; Lyne et al. 2015; Ho et al. 2017.\\
$c$ Casares et al. 2012; Moritani et al. 2018; Malyshev et al. 2019; Adams et al. 2021.\\
$d$ Gregory 2002; Aragona et al. 2009.\\ 
$\dag$ This work.
}}
\end{table*}

\section{Discussion and Conclusion}

Be/$\gamma$-ray binaries provide information as to the relationship between Be stars and high-energy astrophysics (Lamberts 2016; Moritani \& Kawachi 2021). Interaction between these compact objects and their stellar outflow is believed to play an essential role in shaping their multi-wavelength light curves. In this paper, we present our study of the correlation between the keV and TeV emission of Be/$\gamma$-ray binaries under the pulsar scenario. We show that the double-hump structures of keV/TeV flux are caused by the pulsar wind--stellar disc interactions, during which synchrotron and IC radiations are enhanced due to the increases in magnetic field and photon field in the shock. Within this scenario, the disc parameters can be estimated by fitting the light curves.

We found that the keV/TeV light curves of \PSRB1259 and \PSRJ2032 can be well explained by the inclined disc model. Our fitting results suggest that the disc midplane is located at $\phi_{\rm d,\pm}=12^{\circ}\pm 90^{\circ}$ ($\phi_{\rm d,\pm}=165^{\circ}\pm 90^{\circ}$) with an inclination angle of $i_{\rm d}\simeq145^{\circ}$ ($i_{\rm d}\simeq 9^{\circ}$) for LS 2883 (MT91 213).
As for \HESSJ0632 and \LSI61, modelling their orbital modulations with the current available orbital solutions is challenging under the inclined disc model. We used the shapes of light curves and the separations between two peaks to estimate the periastron phases and the orbital eccentricities, and propose alternative orbital geometries.
We suggest that the periastron phase of \HESSJ0632 (\LSI61) is located around $\Phi_{\rm p}\sim 0.35$ ($\Phi_{\rm p}\sim 0.70$ ) with an eccentricity of $e\sim0.35$ ($e\sim0.40$).
We derived the positions and inclination angles of the discs by fitting the keV/TeV light curves (i.e. $i_{\rm d}=12^{\circ}, \phi_{\rm d}=72^{\circ}$ and $i_{\rm d}=10^{\circ}, \phi_{\rm d}=30^{\circ}$ for \HESSJ0632 and \LSI61, respectively). The related observational and model parameters of four Be/$\gamma$-ray binaries are summarised in Table \ref{tab:para}.

Although our model can roughly fit the correlated keV/TeV light curves of observed Be/$\gamma$-ray binaries, the discrepancies between the observational data and model predictions suggest that there are still several limitations to our emission model.
Firstly, we assume that the emitting particles have a mono-energetic distribution, while it is generally believed that the electron injection spectrum follows a power-law function, and the related cooling processes will also reshape the distribution (Khangulyan et al. 2007; Zabalza et al. 2011a; Chen et al. 2019). Therefore, in a more sophisticated model,
the related acceleration and cooling processes of emitting particles should be taken into account.
Secondly, the shock radiations are calculated under a simple one-zone model. It should be noted that, in the realistic case, the emitting region could be much more complicated.
Generally, it is believed that the shock has a bow-shaped geometry wrapping around the pulsar, and the shocked flows moving from the apex to the tail could result in more complicated boosted emissions (Chen et al. 2021b). Also, the hydrodynamic simulations suggested that the pulsar's orbital motion can create another shock on the far side of the binary, which may contribute to observed emissions, in particular for those $\gamma$-ray binaries with more compact orbits (Bosch-Ramon et al. 2012, 2015; Huber et al. 2021a,b; Barkov \& Bosch-Ramon 2021).
Thirdly, we assume that the ram pressure of Be outflow follows a simple $(1+G\cos\theta^m)$ distribution, and therefore that the properties of the disc can be simply characterised by the confinement parameter $m$ and equator-to-pole pressure contrast $G$.
In the widely adopted Gaussian distribution for the disc density, the disc opening angle becomes larger with growing radius (e.g. Carciofi \& Bjorkman 2006, 2008). Using the Gaussian disc model to obtain the shock position requires further assumptions about the disc speeds (including the radial, azimuthal, and vertical directions), which are still quite uncertain for Be stars (e.g. see Sect 4.3 of Torres et al. 2012 for a discussion). Furthermore, if the disc is a Keplerian disc, the rotation of the disc can significantly deflect the shock structure, which has not yet been fully investigated.
The rotation of the disc can significantly change the shock structure, because the disc pressure will push the shock apex much closer to the pulsar if the rotation of the disc is retrograde with the pulsar's orbit, and further away from the pulsar if the
rotation of the disc is prograde. The Keplerian disc can also significantly deflect the tail of the shock from the orbital plane. Depending on the angle between the shock structure and the line of sight, the Doppler boosting effect can also affect the observed flux.
In addition, the inhomogeneity of stellar outflow will cause additional flux variabilities (Bosch-Ramon 2013; Paredes-Fortuny et al. 2015; de la Cita et al. 2017). Also, the anisotropy of the pulsar wind might also affect the radiation (Kong et al. 2012; Bosch-Ramon 2021).
We also ignore the effect of pair absorption on TeV flux, and this process might reduce the TeV photons, especially around periastron and SUPC (Dubus 2006b). The ensuing cascade emissions from the secondary electron--positron pairs will also contribute to the observed emissions at lower energies (e.g. Bosch-Ramon et al. 2008; Takata et al. 2017).

Nevertheless, our simple emission model is still able to capture the effects of Be discs on the shock radiations, and the overall modulations of correlated keV/TeV flux from the four detected Be/$\gamma$-ray binaries are reproduced under the inclined disc model.
The results could be beneficial for future measurements of orbital parameters and searches for potential radio pulsations from compact objects.

\begin{acknowledgements}
We thank Dr. Denys Malyshev for providing us the long-term X-ray data of \HESSJ0632 and the referee for valuable comments on the manuscript. We also want to thank Prof. Kwong-Sang Cheng and Prof. Yun-Wei Yu for useful discussions. This work is supported by the National Key R\&D Program of China (Grant No. 2020YFC2201400), the National Natural Science Foundation of China (Grant No. U1838102), and the China Postdoctoral Science Foundation (Grant No. 2020M682392).
\end{acknowledgements}

\end{document}